\documentclass{article}
\usepackage{spconf,amsmath,graphicx}
\usepackage{spconf,amsmath,graphicx,hyperref}
\usepackage{cite}
\usepackage{amsmath,amssymb,amsfonts}
\usepackage{lscape} 
\usepackage{multirow}  
\usepackage{algorithmic}
\usepackage{graphicx}
\usepackage{textcomp}
\usepackage{amssymb}  
\usepackage{graphicx}  
\usepackage{subcaption}  
\usepackage{xcolor}
\usepackage{soul}
\usepackage{booktabs}

\usepackage{times}
\usepackage{url}
\usepackage{epsfig}
\usepackage{graphicx}
\usepackage{amsmath}
\usepackage{amssymb}
\usepackage{ragged2e}
\usepackage{makecell}
\usepackage[dvipsnames]{xcolor}
\usepackage{tikz}
\usepackage{soul}

\let\oldthebibliography\thebibliography
\let\endoldthebibliography\endthebibliography
\renewenvironment{thebibliography}[1]
{
  \oldthebibliography{#1}
  \setlength{\itemsep}{-0.3pt}
  \setlength{\parsep}{0pt}
  \setlength{\parskip}{0pt}
}
{
  \endoldthebibliography
}

\title{HPGN: Hybrid Priors-Guided Network for Compressed Low-Light Image Enhancement}
\name{Hantang Li~\textsuperscript{\rm 1,2}, Qiang Zhu~\textsuperscript{\rm 2*}, Xiandong Meng~\textsuperscript{\rm 2*}, Lei Xiong~\textsuperscript{\rm 3}, Shuyuan Zhu~\textsuperscript{\rm 3}, Xiaopeng Fan~\textsuperscript{\rm 2,4*}\thanks{* Corresponding Authors.}
}
\address{
\textsuperscript{\rm 1}Harbin Institute of Technology, Shenzhen, \textsuperscript{\rm 2}Pengcheng Laboratory \\
\textsuperscript{\rm 3}University of Electronic Science and Technology of China, \textsuperscript{\rm 4}Harbin Institute of Technology \\
\{25B951062@stu., fxp@\}hit.edu.cn, \{zhuqiang, mengxd\}@pcl.ac.cn \\
\{leixiong@std., eezsy@\}uestc.edu.cn
}

\begin{document}
%
\maketitle
\begin{abstract}
In practical applications, low-light images are often compressed for efficient storage and transmission. Most existing methods disregard compression artifacts removal or hardly establish a unified framework for joint task enhancement of low-light images with varying compression qualities. 
To address this problem, we propose an efficient hybrid priors-guided network (HPGN) that enhances compressed low-light images by integrating both compression and illumination priors. Our approach fully utilizes the JPEG quality factor (QF) and DCT quantization matrix (QM) to guide the design of efficient plug-and-play modules for joint tasks. 
Additionally, we employ a random QF generation strategy to guide model training, enabling a single model to enhance low-light images with different compression levels. Experimental results demonstrate the superiority of our proposed method.
\end{abstract}
\section{Introduction}
\label{sec:intro}

Low-light images are captured under challenging lighting conditions using conventional equipment, often exhibiting poor visual quality and negatively impacting other low-level vision~\cite{zhang2017beyond,zhu2023attention,zhu2025blind} and high-level vision tasks~\cite{zerodce,snraware} such as denoising, super-resolution, object detection, and tracking. Moreover, in practical applications, low-light images often require compression to reduce storage and transmission costs, which can result in compression artifacts and reduced quality of low-light images~\cite{capformer}.

\begin{figure}[!t]
\centering
\includegraphics[width=0.49\textwidth]{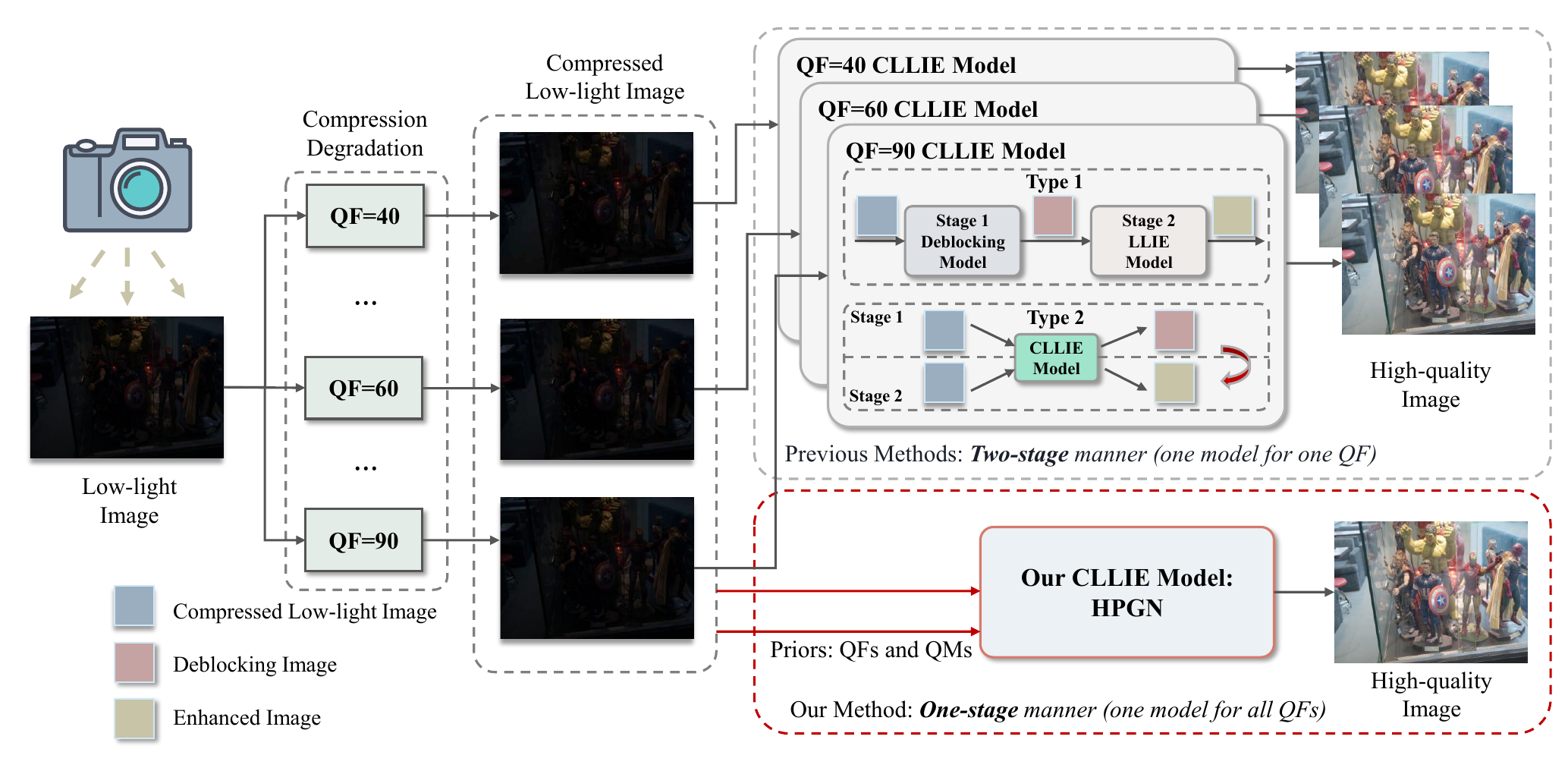}
\caption{Comparison between previous compressed low-light image enhancement (CLLIE) methods \cite{capformer,fbcnn,mirnetv2,fourllie,retinexformer} and our HPGN. Our HPGN model is capable of simultaneously improving compressed low-light images of varying qualities without requiring repeated model training.}
\label{fig:framecom}
\end{figure}

Although many existing low-light image enhancement (LLIE) methods \cite{zerodce, enlightengan, Kind, restormer, snraware, zhu2024temporally, pairllie,ciftci2025lowlight} have achieved significant brightness adjustment, they do not effectively mitigate compression artifacts when applied to compressed low-light images. Some methods adopt a two-stage manner to address this problem. For example, cascading an image deblocking model\cite{fbcnn,zhu2024cpga,dualstream_jpeg,mo2024oapt,zhu2026fcvsr,meng2020robust,zhu2024compressed,zhu2024deep,meng2020flow} and an LLIE model\cite{fourllie,snraware,hvi,retinexformer,darkir,pairllie} for compressed low-light image enhancement (CLLIE). Moreover, CAPformer \cite{capformer} first trained a model on compressed and uncompressed low-light image pairs, then fine-tuned the pre-trained model on compressed low-light images and uncompressed high-quality image pairs to generate enhanced images.
These two-stage methods can optimize certain details, but they often lead to significant noise distortion. Besides, models trained for a specific QF are typically evaluated only at that same QF. The training time costs will increase by multiples for different QFs. 
As illustrated in Fig.~\ref{fig:framecom}, we account for compression and illumination characteristics to develop a one-stage compressed low-light enhancement model to deliver robust enhancement only training once across images with varying compression qualities. 

\begin{figure*}[!t]
\centering
\includegraphics[width=1.00\textwidth]{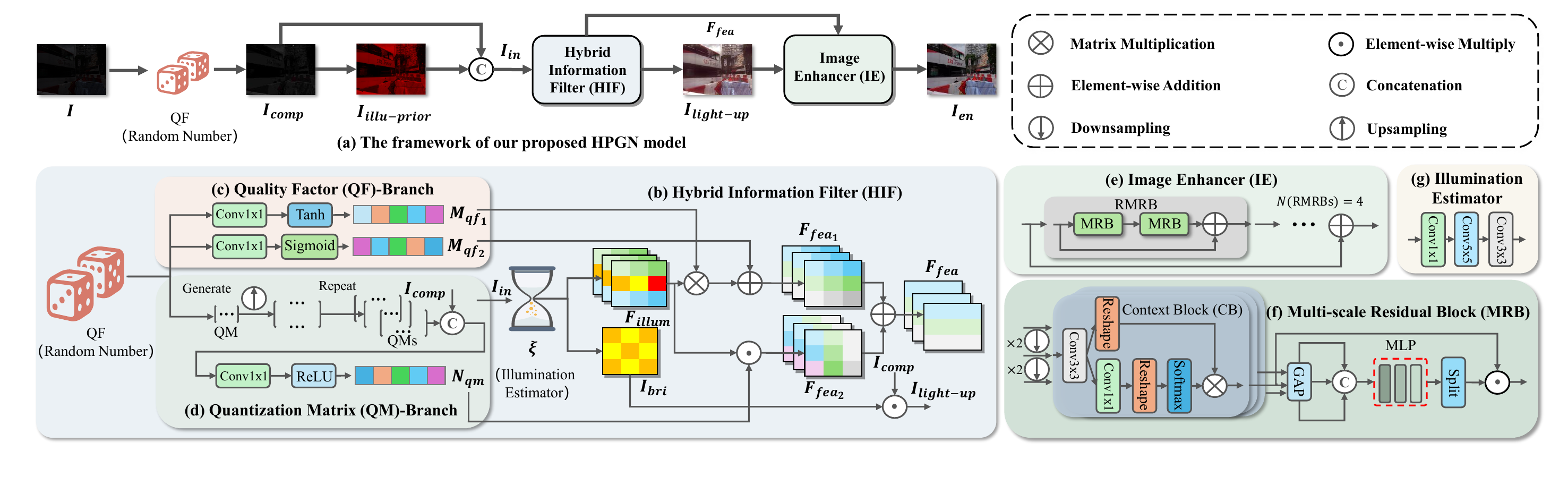} 
\caption{(a) The framework of our HPGN model. (b)-(d) The hybrid information filter (HIF) integrates illumination and compression priors to dynamically adjust features for improving the image quality. (e) The image enhancer (IE) consists of stacked recursive multi-scale residual blocks (RMRB), each containing several (f)  multi-scale residual blocks (MRB). (g) The Illumination Estimator is used for local illumination feature extraction and coarse enhancement of compressed low-light images.}
\label{fig:label} 
\end{figure*}

In this paper, we design a plug-and-play hybrid information filter (HIF) that combines the illumination prior with two compression priors, i.e., quality factor (QF) and the discrete cosine transform (DCT) \cite{dct} quantization matrix (QM). 
This filter effectively eliminates compressed artifacts and improves the quality of illumination features during the enhancement process. Additionally, a random QF generation strategy is employed during the training, enabling the model to robustly enhance compressed low-light images across varying QFs, thereby reducing computational resources and training time.

The main contributions are listed as follows:
\begin{itemize}
\item We propose a one-stage compressed low-light image enhancement network, dubbed as HPGN, guided by prior knowledge of illumination and compression.

\item Hybrid information filter (HIF) was designed that leverages compression priors, i.e., QF, QM, with illumination prior to mitigate the information loss caused by image compression during brightness adjustment. Moreover, the plug-and-play advantage of HIF allows it to be embedded in other low-light enhancement methods. 

\item A random QF generation strategy is employed during model training, enabling a single model to effectively enhance compressed low-light images across varying QFs. This strategy reduces training time costs compared to the other two-stage methods. 
\end{itemize}

Experimental results have demonstrated that the proposed HPGN has state-of-the-art performance with four methods on four datasets. Our HPGN outperforms the second-best approach, CAPformer~\cite{capformer}, by up to 0.21 dB in PSNR and reduces 69\% of model parameters for low-light image enhancement with varying levels of compression, only training once.

\section{Related Work}

\subsection{Low-Light Image Enhancement}

Low-light image enhancement (LLIE) has attracted significant attention in recent years. Early learning-based approaches typically adopt end-to-end frameworks to directly learn a mapping from low-light images to normal-light images, achieving notable improvements in brightness and visual quality~\cite{zerodce,enlightengan,snraware,fourllie,pairllie}. Another prominent line of research is inspired by Retinex theory, which decomposes an image into illumination and reflectance components and enhances low-light images by estimating and correcting the illumination map~\cite{Kind,retinexformer,darkir}. While these methods achieve strong performance on uncompressed low-light images, they are not explicitly designed to handle compression artifacts and often suffer from noise amplification and detail distortion when applied to compressed low-light images.

\subsection{Compressed Low-Light Image Enhancement}

Enhancing compressed low-light images is particularly challenging due to the compounded degradation from low illumination and compression artifacts. A common strategy is a two-stage enhancement, in which enhancement and artifact removal are cascaded, often leading to error accumulation and increased model complexity. More recently, CAPformer~\cite{capformer} proposed a compression-aware framework by fine-tuning a pre-trained model for specific JPEG quality factors. However, this approach still requires separate optimization for each compression level and lacks flexibility in handling images with varying qualities. These limitations motivate the development of a unified and efficient one-stage framework for compressed low-light image enhancement.

\section{Method}

The overall framework of our hybrid priors-guided compressed low-light image enhancement network is shown in Fig.~\ref{fig:label} (a), which includes a hybrid information filter (HIF) module (in Fig.~\ref{fig:label} (b)-(d), (g)) and a CNN-based image enhancer (IE) module (in Fig.~\ref{fig:label} (e)-(f)). During model training, the quality factor (QF) is randomly generated based on the input low-light image $I$ and subjected to specific JPEG compression. The generated QF is then input into the HIF for quantization matrix (QM) generation and illumination feature adjustment. The adjusted features are subsequently passed to the IE to assist in obtaining the final enhanced image $I_{en}$. 

\subsection{Hybrid Information Filter}

To effectively utilize the characteristics of illumination and compression to achieve the enhancement of compressed low-light images, we design a hybrid information filter (HIF) to dynamically adjust features based on two characteristics, i.e., QF and QM, to improve the image enhancement results. The HIF module consists of an illumination estimator, a QF-branch, and a QM-branch (as shown in Fig.~\ref{fig:label}). 

Based on existing LLIE methods \cite{retinexformer, retinexmamba, iagc}, we perform global brightness map estimation and local illumination feature extraction \cite{retinexformer} from compressed low-light image $I_{comp}$:
\begin{align}
I_{illu\text{-}prior} &= \delta (I_{comp}), \\
I_{bri}, F_{illum} &= \xi(I_{comp}, I_{illu\text{-}prior}), \notag \\
I_{light\text{-}up} &= I_{bri} \odot I_{comp},\label{eq:formula1}
\end{align}
where $\delta(\cdot)$ represents the operation of calculating the average value of each pixel of the image along the channel dimension to generate an illumination prior map $I_{illu\text{-}prior}$. An illumination estimator $\xi(\cdot)$ shown in Fig.~\ref{fig:label} (g). It outputs the global brightness estimation map $I_{bri}$ and local illumination features {$F_{illum}$}. $I_{bri}$ is then multiplied with $I_{comp}$ to obtain a preliminary enhanced image $I_{light\text{-}up}$. 

For a compressed image, the importance distribution of each channel feature may change (e.g., color information loss or edge blurring) after compression, and the QF is directly related to the correlation of the compressed channels in the image. Therefore, dynamically assigning different weights to each channel allows for more accurate enhancement or suppression of important channels, thereby alleviating feature shift caused by quantization. We propose a QF-branch in the HIF for this idea, and its calculation expression is:
\begin{align}
    F_{fea_1} &= F_{illum} \otimes  M_{{qf_1}} + M_{{qf_2}}, \label{eq:formula2}
\end{align}
where $M_{qf_1}$, $ M_{qf_2} $are the mapping coefficients related to QF and $\otimes$ denotes the channel-wise multiplication.

QM also encodes the importance of details at different positions in the image (such as texture or edge preservation) and can directly influence the spatial feature distribution during enhancement. 
Therefore, we design a QM-branch in the HIF to integrate QM into the illumination features, effectively modeling spatial features during the enhancement process, amplifying the enhancement effect in key areas through attention mechanisms, and locally enhancing regions with significant quantization degradation. Specifically, since the QM is defined on $8 \times 8$ DCT blocks, it is repeated across non-overlapping $8 \times 8$ patches to match the feature map resolution, preserving alignment with the JPEG grid. Its calculation expression is:
\begin{align}
    F_{fea_2} &= F_{illum}  \odot N_{qm},\label{eq:formula3}
\end{align}
where $N_{qm}$ is the mapping coefficients related to QM and $ \odot$  represents spatial element-wise multiplication.

The final calculation expression $ F_{fea}$ for the output features combining illumination and compression priors is:
\begin{align}
    F_{fea} &= F_{fea_1} + F_{fea_2}.\label{eq:formula4}
\end{align}

HIF is essentially a plug-and-play feature filtering control module with excellent scalability and adaptability. This module can be integrated with existing LLIE methods to enhance their performance in processing compressed low-light image enhancement tasks, as verified in the experimental section.

\subsection{Image Enhancer}

To achieve the high-quality enhancement performance while low complexity, we design an Image Enhancer (IE) consisting of multiple recursive multi-scale residual blocks (RMRB) stacked together. Each recursive residual block contains several multi-scale residual blocks (MRB). The structure of IE is shown in Fig.~\ref{fig:label} (e). 
We adopt a multi-scale branch architecture in MRB, where the input feature map is downsampled twice to generate feature maps at different resolutions to overcome the varying light distribution scale.  
Each branch is extracted using the context block (CB) to focus on capturing the correlation between local and global features. Then, a dynamic weighted fusion of information is performed from different scales. 
Finally, the input features are adaptively fused at the channel level and combined with the residual connection to generate the final output. This approach ensures more effective feature integration and significantly enhances the model's overall representational capacity. 

\subsection{Loss Function}

We use the  L1 loss and perceptual loss to train our model. Our loss function is defined as follows:
\begin{equation}
\mathcal{L}=\left\|I_{en}-I_{high}\right\|_1+\lambda_{per}\left\|\phi\left(I_{en}\right)-\phi\left(I_{high}\right)\right\|_1,
\end{equation}
where $I_{en}$ and  $I_{high}$ represent the enhanced image and the corresponding ground truth image. $\phi(\cdot)$ denotes the pre-trained VGG19 network. We set the weight $\lambda_{per}$ to 0.01.

\begin{table*}[!t]
\setlength{\tabcolsep}{0.800mm}
\renewcommand\arraystretch{1}
\fontsize{9}{9}\selectfont
\centering
\caption{Comparison results at QF=80 on LOLv1, LOLv2-real, and LOLv2-syn datasets and at random QF on LOLv1 dataset.}
\label{table:comparison_methods_2}
\begin{tabular}{c|c|c c|c c|c c||c c} 
\toprule
\multirow{2}{*}{\text{Methods}} & \multirow{2}{*}{\text{Params. (M)}} & \multicolumn{2}{c|}{\text{LOLv1 (QF=80)}} & \multicolumn{2}{c|}{\text{LOLv2-real (QF=80)}} & \multicolumn{2}{c||}{\text{LOLv2-syn (QF=80)}} & \multicolumn{2}{c}{\text{LOLv1-randomQF}} \\ 
& & \text{PSNR (dB)$\uparrow$} & \text{SSIM}$\uparrow$ & \text{PSNR (dB)$\uparrow$} & \text{SSIM}$\uparrow$ & \text{PSNR (dB)$\uparrow$} & \text{SSIM}$\uparrow$ & \text{PSNR (dB)$\uparrow$} & \text{SSIM}$\uparrow$ \\
\midrule
MIRNet\cite{mirnet} (LLIE)& 31.76 & 21.315 & 0.777 & 20.447 & 0.768 & 22.417 & 0.826 & 21.238 & 0.741\\
MIRNetv2\cite{mirnetv2}(LLIE) & 5.9 & 22.343 & 0.792 & 21.482 & 0.783 & 22.793 & 0.832 & 21.784 & 0.772 \\
PairLIE\cite{pairllie} (LLIE)& 0.33 & 18.087 & 0.623 & 17.854 & 0.582 & 20.654 & 0.705 & 17.845 & 0.612\\
FourLLIE\cite{fourllie} (LLIE)& 0.12 & 20.644 & 0.745 & 19.902 & 0.749 & 21.737 & 0.801 & 20.123 & 0.721\\
Retinexformer\cite{retinexformer} (LLIE)& 1.61 & 22.757 & 0.779 & 21.064 & 0.773 & 22.585 & 0.826 & 22.323 & 0.747 \\
HVI-CIDNet\cite{hvi} (LLIE)& 1.88 & 22.936 & 0.790 & 21.156 & 0.782 & 22.644 & 0.828 & 22.469 & 0.754 \\
FBCNN\cite{fbcnn} + MIRNetv2\cite{mirnetv2}& 71.92+5.90 & 22.597 & 0.801 & 21.564 & 0.794 & 23.021 & 0.817 & 21.965 & 0.765 \\
FBCNN\cite{fbcnn} + FourLLIE\cite{fourllie} & 71.92+0.12 & 21.001 & 0.772 & 20.436 & 0.764 & 21.981 & 0.811 & 20.542 & 0.733 \\
FBCNN\cite{fbcnn} + Retinexformer\cite{retinexformer} & 71.92+1.61 & 22.873 & 0.793 & 20.865 & 0.787 & 23.150 & 0.838 & 22.123 & 0.795 \\
FBCNN\cite{fbcnn} + HVI-CIDNet\cite{hvi} & 71.92+1.88 & 23.146 & 0.801 & 21.276 &  0.791 & 23.194 & 0.837 & 22.472& 0.815 \\
CAPformer\cite{capformer} & 8.77 & \textbf{23.499} & 0.807 & 21.689 & 0.797 & 23.296 & 0.840 & 22.634 & 0.783  \\
\midrule
\textbf{HPGN (Ours)} & \textbf{2.69} & 23.333 & \textbf{0.833} & \textbf{21.924} & \textbf{0.844} & \textbf{23.443} & \textbf{0.875} & \textbf{22.844} & \textbf{0.826}  \\
\bottomrule
\end{tabular}
\end{table*}

\section{Experiments}

\subsection{Datasets and Experimental Settings}

We use LOLv1\cite{retinexnet}, LOLv2-real\cite{lolv2}, and LOLv2-syn\cite{lolv2} as our benchmark datasets. We randomly control the QF parameters of JPEG compression for both the training and test datasets. To evaluate the effectiveness of our proposed model, we randomly generate QF on the LOLv1 dataset and applied specified JPEG compression based on these values, resulting in a new dataset called LOLv1-randomQF.
As recommended in \cite{capformer}, JPEG-compressed images with a QF of 80 offer a good balance between storage efficiency and visual quality; thus, QF=80 is adopted in the experiment. The number of RMRBs and MRBs is set to 4 and 2. All experiments are conducted on a single NVIDIA GeForce RTX 2080Ti GPU using PyTorch. For training hyperparameters (e.g., optimizer, learning rate, and batch size), we follow the exact settings of MIRNetv2~\cite{mirnetv2} to ensure optimal convergence.

For fairness, we retrain the models on the JPEG-compressed low-light dataset for both single low-light and compressed image enhancement tasks. 
Additionally, since this is a joint task, we include a comparison with a cascading decompression and low-light enhancement method. 
Performance of models is evaluated using the PSNR and SSIM metrics. 

\subsection{Quantitative Evaluation}


We first evaluate the performance of the proposed model in processing JPEG compressed images with a single QF value; the experimental results on three datasets with QF=80 are shown in Table~\ref{table:comparison_methods_2}. It can be found that our method achieves optimal performance on both the LOLv2-real and LOLv2-syn datasets. Additionally, our method ranks second on the LOLv1 dataset. Notably, compared to the SOTA method, CAPformer on the LOLv1 dataset, our approach contains only 30.7\% of its model parameters.

To further evaluate the performance of the proposed model in processing JPEG compressed images with random QF, we perform the experiment on the LOLv1-randomQF dataset and the results are presented in Table~\ref{table:comparison_methods_2}. As shown, our proposed method outperforms LLIE methods (MIRNet\cite{mirnet}, MIRNetv2\cite{mirnetv2}, PairLIE\cite{pairllie}, FourLLIE\cite{fourllie}, Retinexformer\cite{retinexformer}, HVI-CIDNet\cite{hvi}) and two-stage methods (FBCNN\cite{fbcnn} + MIRNetv2\cite{mirnetv2}, FBCNN\cite{fbcnn} + PairLIE\cite{pairllie}, FBCNN\cite{fbcnn} + Retinexformer\cite{retinexformer}, FBCNN\cite{fbcnn} + HVI-CIDNet\cite{hvi}, CAPformer\cite{capformer}), achieving the highest results in both PSNR and SSIM. 
Compared to LLIE methods, although two-stage methods offer some improvements for joint tasks, they introduce a large number of parameters, significantly increasing the model's complexity. 
Furthermore, our method incorporates JPEG-related parameters such as QF and QM, along with advanced training strategies, setting it apart from existing SOTA joint task processing methods. 
This enables flexible enhancement of JPEG compressed images at various compression levels, demonstrating the versatility and superiority of our approach.

To demonstrate the scalability of our proposed core module, i.e., HIF, Table~\ref{table:plug-and-play} presents the experimental results of its integration with two advanced LLIE frameworks, i.e., MIRNetv2\cite{mirnetv2} and Retinexformer\cite{retinexformer}, on the LOLv1-randomQF dataset.  It is found that even with the addition of a small number of parameters, our HIF significantly improves the performance of these methods in handling joint tasks.

\begin{figure*}[!t]
\centering

\begin{subfigure}{0.24\textwidth}
\centering
\includegraphics[width=\linewidth]{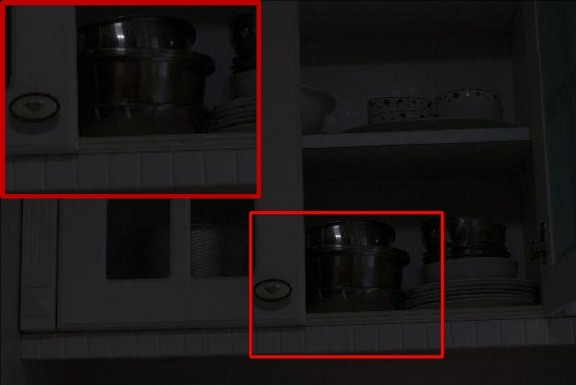}
\footnotesize Input 1
\end{subfigure}
\begin{subfigure}{0.24\textwidth}
\centering
\includegraphics[width=\linewidth]{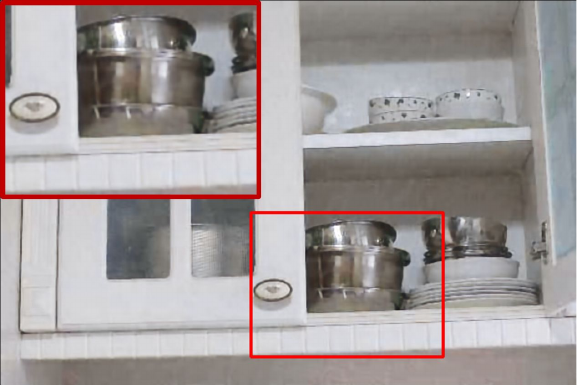}
\footnotesize Retinexformer
\end{subfigure}
\begin{subfigure}{0.24\textwidth}
\centering
\includegraphics[width=\linewidth]{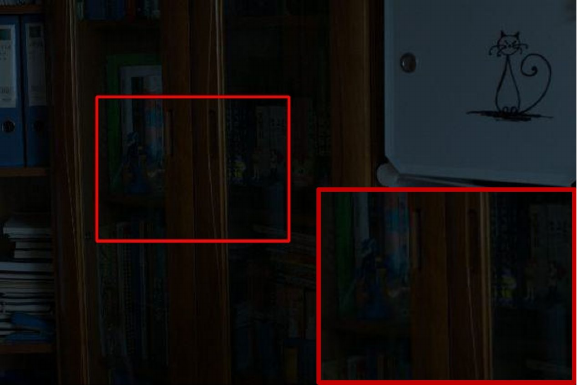}
\footnotesize Input 2
\end{subfigure}
\begin{subfigure}{0.24\textwidth}
\centering
\includegraphics[width=\linewidth]{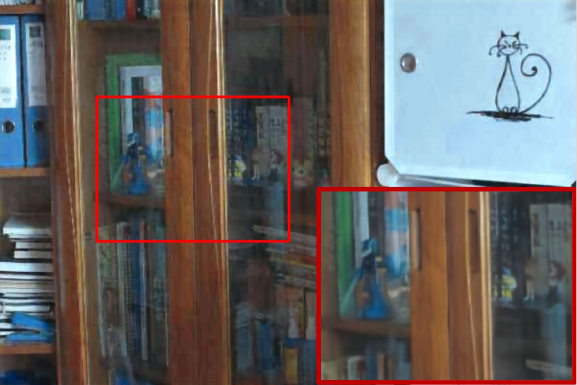}
\footnotesize Retinexformer
\end{subfigure}

\vspace{2pt}

\begin{subfigure}{0.24\textwidth}
\centering
\includegraphics[width=\linewidth]{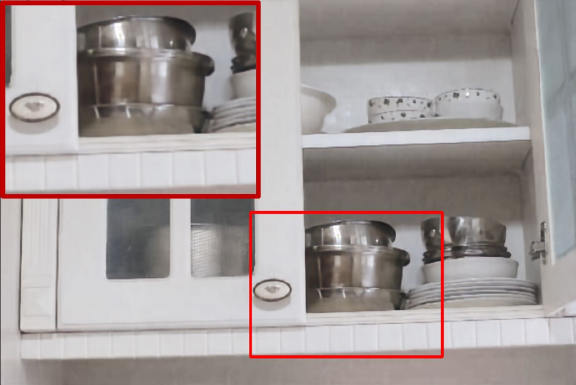}
\footnotesize FBCNN+Retinexformer
\end{subfigure}
\begin{subfigure}{0.24\textwidth}
\centering
\includegraphics[width=\linewidth]{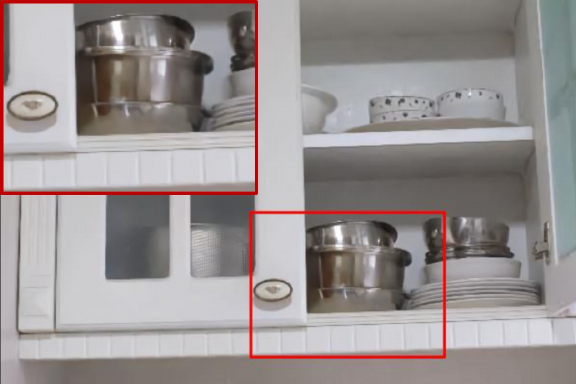}
\footnotesize CAPformer
\end{subfigure}
\begin{subfigure}{0.24\textwidth}
\centering
\includegraphics[width=\linewidth]{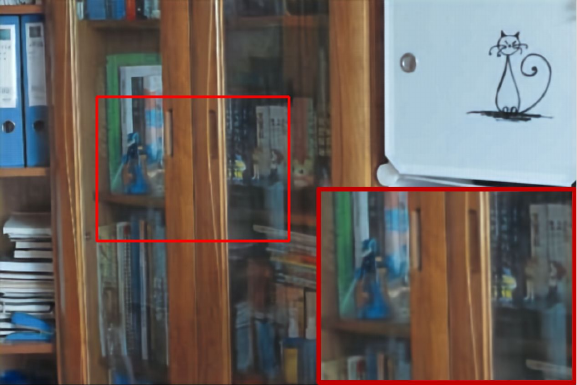}
\footnotesize FBCNN+Retinexformer
\end{subfigure}
\begin{subfigure}{0.24\textwidth}
\centering
\includegraphics[width=\linewidth]{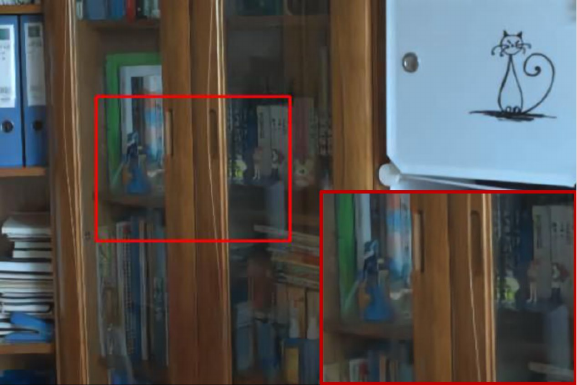}
\footnotesize CAPformer
\end{subfigure}

\vspace{2pt}

\begin{subfigure}{0.24\textwidth}
\centering
\includegraphics[width=\linewidth]{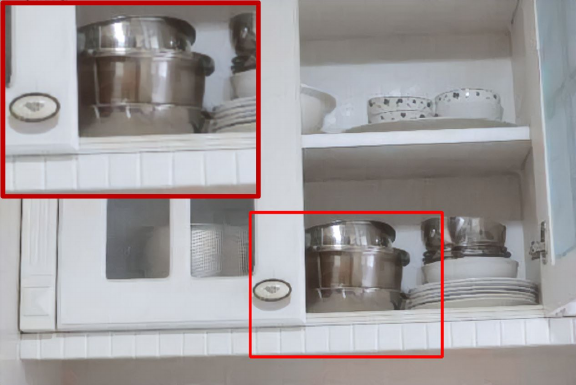}
\footnotesize HPGN (Ours)
\end{subfigure}
\begin{subfigure}{0.24\textwidth}
\centering
\includegraphics[width=\linewidth]{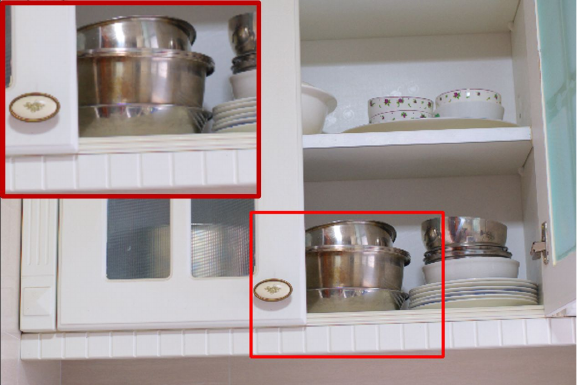}
\footnotesize GT
\end{subfigure}
\begin{subfigure}{0.24\textwidth}
\centering
\includegraphics[width=\linewidth]{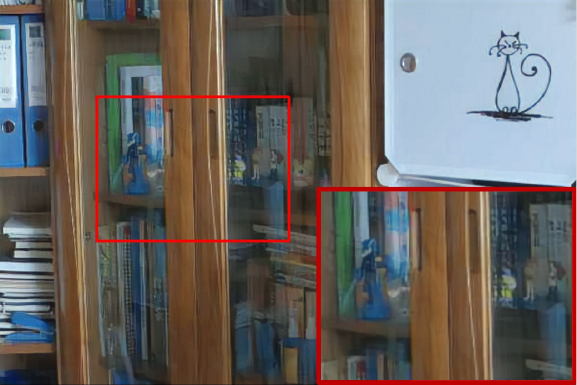}
\footnotesize HPGN (Ours)
\end{subfigure}
\begin{subfigure}{0.24\textwidth}
\centering
\includegraphics[width=\linewidth]{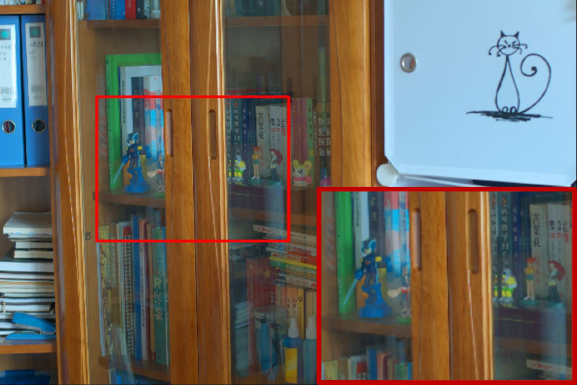}
\footnotesize GT
\end{subfigure}

\caption{Visual comparison on the LOLv1-randomQF dataset.}
\label{fig:example}
\vspace{-6pt}
\end{figure*}

\subsection{Qualitative Evaluation}

We present the visual results of our proposed method against  ‌representative‌ methods, as well as an LLIE method (Retinexformer\cite{retinexformer}), two-stage methods (FBCNN\cite{fbcnn} + Retinexformer\cite{retinexformer}, CAPformer\cite{capformer}) on the LOLv1-randomQF dataset in Fig.~\ref{fig:example}. 
Although the LLIE methods have been retrained on the compressed low-light dataset, its enhancement results still fail to effectively prevent compression artifacts and color distortion. While two-stage methods have alleviated these issues to some extent, they remain inferior to our method and are more complex. 
Our results demonstrate superior visual quality and reduced noise, indicating improved performance.

\subsection{Ablation Study}

In this section, we conduct three ablation experiments by individually removing different branches from the HIF. The evaluation is performed on the LOLv1-randomQF dataset. Table~\ref{table:ablation} presents the results of the ablation experiments, where the baseline model was processed using only the IE for joint tasks. As shown in Table~\ref{table:ablation}, incorporating the QF and QM branches into HIF significantly improves the baseline model's performance in handling joint tasks.

\begin{table}[!t]
\centering
\setlength{\tabcolsep}{0.800mm}
\renewcommand\arraystretch{1}
\fontsize{9}{9}\selectfont
\caption{HIF validation on LOLv1-randomQF dataset.}
\label{table:plug-and-play}
\resizebox{\columnwidth}{!}{ 
\begin{tabular}{c|c|c c} 
\toprule
\multirow{2}{*}{\text{Methods}} & \multirow{2}{*}{\text{Params. (M)}} & \multicolumn{2}{c}{\text{LOLv1-randomQF}} \\ 
& & \text{PSNR (dB)$\uparrow$} & \text{SSIM}$\uparrow$ \\
\midrule
MIRNetv2\cite{mirnetv2} & 5.90 & 21.784 & 0.772 \\
MIRNetv2\cite{mirnetv2} + \textbf{HIF (Ours)}  & 6.33 & \textbf{22.124} & \textbf{0.784} \\
\midrule
Retinexformer\cite{retinexformer} & 1.61 & 22.323 & 0.747 \\
Retinexformer\cite{retinexformer} + \textbf{HIF (Ours)} & 1.68 & \textbf{22.547} & \textbf{0.762} \\
\bottomrule
\end{tabular}
}
\end{table}

\begin{table}[!t]
\centering
\setlength{\tabcolsep}{2.00mm}
\renewcommand\arraystretch{1}
\fontsize{9}{9}\selectfont
\caption{Ablation studies on LOLv1-randomQF dataset. }
\label{table:ablation}
\begin{tabular}{c c c|c c} 
\toprule
\multirow{2}{*}{\text{Baseline}} & \multirow{2}{*}{\text{QF-branch}} & \multirow{2}{*}{\text{QM-branch}} & \multicolumn{2}{c}{\text{LOLv1-randomQF}} \\ 
& & & \text{PSNR (dB)$\uparrow$} & \text{SSIM}$\uparrow$ \\
\midrule  
\checkmark &  & & 22.044 & 0.797 \\
\checkmark & \checkmark & & 22.532 & 0.824 \\
\checkmark & & \checkmark & 22.405 & 0.812 \\
\midrule 
\checkmark & \checkmark & \checkmark & \textbf{22.844} & \textbf{0.826} \\
\bottomrule 
\end{tabular}
\vspace{-5pt}
\end{table}

\vspace{-5pt}

\section{Discussion and Future Work}

Although our experiments mainly focus on the widely used JPEG standard, the core idea of HPGN uses compression-related metadata to guide image enhancement and is not limited to a specific compression artifact. Extending it to other compression standards or extremely severe compression with QF $< 10$ would require customized prior modeling and adaptations. A systematic study of these scenarios is beyond the scope of this paper due to space limitations. Nevertheless, HPGN can serve as an extensible baseline for these challenging scenarios, which we plan to investigate in future work.

\section{Conclusion}

In this paper, we propose a one-stage compressed low-light image enhancement method called hybrid priors-guided network (HPGN), which leverages illumination and compression priors. By integrating the quality factor (QF) of encoding and the DCT quantization matrix (QM) with the illumination prior in our hybrid information filter, our method effectively addresses the challenges of both JPEG compression and low-light enhancement. Additionally, the random QF generation strategy facilitates the training of a single model across various compression levels, enabling it to enhance compressed low-light images of varying qualities while reducing training time costs. Furthermore, the hybrid information filter (HIF) boosts the performance of existing enhancement methods and can be seamlessly integrated as a plug-and-play solution.
Experimental results show that our method outperforms existing approaches in compressed low-light image enhancement across different compression levels.

\clearpage

\begingroup
\setlength{\itemsep}{0pt}
\setlength{\parskip}{0pt}

\bibliographystyle{IEEEbib}

\bibliography{strings,refs}
\endgroup
\end{document}